\documentclass[10pt,journal,compsoc]{IEEEtran}

\usepackage{amsfonts,mathrsfs,latexsym,amsmath,amssymb,amsthm}

\ifCLASSOPTIONcompsoc
  \usepackage[nocompress]{cite}
\else
  \usepackage{cite}
\fi
\ifCLASSINFOpdf
\else
\fi

\usepackage{amsmath}
\usepackage{caption}
\usepackage{tikz}
\usetikzlibrary{arrows}
\makeatletter
\pgfdeclareshape{datastore}{
  \inheritsavedanchors[from=rectangle]
  \inheritanchorborder[from=rectangle]
  \inheritanchor[from=rectangle]{center}
  \inheritanchor[from=rectangle]{base}
  \inheritanchor[from=rectangle]{north}
  \inheritanchor[from=rectangle]{north east}
  \inheritanchor[from=rectangle]{east}
  \inheritanchor[from=rectangle]{south east}
  \inheritanchor[from=rectangle]{south}
  \inheritanchor[from=rectangle]{south west}
  \inheritanchor[from=rectangle]{west}
  \inheritanchor[from=rectangle]{north west}
  \backgroundpath{
    \southwest \pgf@xa=\pgf@x \pgf@ya=\pgf@y
    \northeast \pgf@xb=\pgf@x \pgf@yb=\pgf@y
    \pgfpathmoveto{\pgfpoint{\pgf@xa}{\pgf@ya}}
    \pgfpathlineto{\pgfpoint{\pgf@xb}{\pgf@ya}}
    \pgfpathmoveto{\pgfpoint{\pgf@xa}{\pgf@yb}}
    \pgfpathlineto{\pgfpoint{\pgf@xb}{\pgf@yb}}
 }
}
\makeatother

\usepackage{graphicx}
\usepackage{array}
\usepackage{multirow}
\usepackage{longtable}
\usepackage{rotating}

\usepackage{supertabular}
\usepackage{multirow}
\usepackage{makecell}
\usepackage{longtable}

\begin{document}
%
\title{New $(k,l,m)$-verifiable multi-secret sharing  \\ schemes based on XTR public key system}
%
%
%
%

\author{Jing~Yang~
        and~Fang-Wei~Fu
\IEEEcompsocitemizethanks{\IEEEcompsocthanksitem Jing Yang is with the Chern Institute of Mathematics and LPMC, and Tianjin Key Laboratory of Network and Data Security Technology, Nankai University, Tianjin, 300071, China.\protect\\
E-mail: yangjing0804@mail.nankai.edu.cn
\IEEEcompsocthanksitem Fang-Wei~Fu is with the Chern Institute of Mathematics and LPMC, and Tianjin Key Laboratory of Network and Data Security Technology, Nankai University, Tianjin, 300071, China.\protect\\
E-mail: fwfu@nankai.edu.cn}
\thanks{Manuscript received April 19, 2015; revised August 16, 2015. (Corresponding author: Jing Yang.)}}

%
%

\markboth{Journal of \LaTeX\ Class Files,~Vol.~14, No.~8, August~2015}%
{Shell \MakeLowercase{\textit{et al.}}: Bare Advanced Demo of IEEEtran.cls for IEEE Computer Society Journals}
%



\IEEEtitleabstractindextext{%
\begin{abstract}
Secret sharing was proposed primarily in 1979 to solve the problem of key distribution. In recent decades, researchers have proposed many improvement schemes. Among all these schemes, the verifiable multi-secret sharing (VMSS) schemes are studied sufficiently, which share multiple secrets simultaneously and perceive malicious dealer as well as participants. By pointing out that the schemes presented by Dehkordi and Mashhadi in 2008 cannot detect some vicious behaviors of the dealer, we propose two new VMSS schemes by adding validity check in the verification phase to overcome this drawback. Our new schemes are based on XTR public key system, and can realize $GF(p^{6})$ security by computations in $GF(p^{2})$ without explicit constructions of $GF(p^{6})$, where $p$ is a prime. Compared with the VMSS schemes using RSA and linear feedback shift register (LFSR) public key cryptosystems, our schemes can achieve the same security level with shorter parameters by using trace function. What's more, our schemes are much simpler to operate than those schemes based on Elliptic Curve Cryptography (ECC). In addition, our schemes are dynamic and threshold changeable, which means that it is efficient to implement our schemes according to the actual situation when participants, secrets or the threshold needs to be changed.
\end{abstract}

}

\maketitle

\IEEEdisplaynontitleabstractindextext

%
\IEEEpeerreviewmaketitle

\ifCLASSOPTIONcompsoc
\IEEEraisesectionheading{\section{Introduction}\label{sec:introduction}}
\else
\section{Introduction}
\label{sec:introduction}
\fi

%
%
%
%
\IEEEPARstart{I}{t} is well-known that secret sharing schemes\cite{Shamir,Blakley} are significant to protect secret keys, which are critical components in many applications of modern cryptography, such as threshold cryptography \cite{DF}, commitment scheme \cite{BN}, secure multiparty computation \cite{CDM,C}, blockchain \cite{KRKV} and so on.

In 2004, an efficient VMSS scheme was proposed by Yang et al. \cite{YCH}. Based on this scheme, in 2005, Shao and Cao \cite{SC} presented an improved scheme. However, this scheme still requires a private channel. In 2006, Zhao et al. (ZZZ) \cite{ZZZ} proposed a new VMSS scheme, where participants choose their own shadows by themselves so that this scheme does not need a security channel. Then, in 2007, Dehkordi and Mashhadi \cite{DM0} introduced RSA public key cryptosystem \cite{RSA} into VMSS schemes for the first time to make their verifiable property more efficient. However, all these schemes \cite{YCH,SC,ZZZ,DM0} still use Lagrange interpolation to distribute secrets, which are similar to Shamir's secret sharing scheme \cite{Shamir}.

Further, in 2008, Dehkordi and Mashhadi presented two new types of efficient VMSS schemes (DM1 \cite{DM1} and DM2 \cite{DM2}), which employ the homogeneous and nonhomogeneous linear recursions \cite{NLB} to increase the efficiency of the construction and reconstruction phase, respectively. In order to reduce the operating time, Hu et al. (HLC) \cite{HLC} utilized LFSR sequence and LFSR public key cryptosystem \cite{GH1,GH2} to verify the validity of the data. Then, in 2015, Dehkordi and Mashhadi (DM3) \cite{DM3} used LFSR public key cryptosystem and new nonhomogeneous linear recursions to make their schemes have shorter private and public key length.

Nevertheless, Liu et al. (LZZ) \cite{LZZ} found that ZZZ and DM1 schemes cannot detect some dealer's hostile behaviors and presented new schemes by RSA public key cryptosystem. Similarly, DM2 \cite{DM2} and DM3 \cite{DM3} have the same drawback as mentioned in \cite{LZZ}. We have proposed modified schemes (YF) \cite{YF} based on DM3 schemes by using LFSR public key cryptosystem. YF schemes can not only perceive the deception of both participants and the dealer, but also use one-third of the private and public key length of LZZ to achieve the same security level.

In this work, we propose two novel VMSS schemes to improve DM2 schemes by XTR public key system \cite{XTR}, which make full use of trace function to reduce the storage of data, computation cost and communication cost. In fact, XTR public key system can realize the security level in $GF(p^{6})$ by computations in $GF(p^{2})$ where $p$ is a prime. Compared with RSA, LFSR public key cryptosystems and ECC, XTR public key system needs shorter key length than RSA and LFSR public cryptosystems to achieve the same security level, and has simpler procedure of parameter and key generation than ECC. Further, XTR public key system can be considered as a special case of the optimization of LFSR public key cryptosystem . Therefore, our proposed VMSS schemes have many good properties which will be discussed later.

The rest of this paper is organized as follows. In Section 2, we review the nonhomogeneous linear recursion, XTR public key system, and give the review and attack on DM2 schemes. In Sections 3 and 4, we propose our two new VMSS schemes respectively. We present the security analysis in Section 5, and in Section 6 we give the performance analysis. Finally, we conclude our schemes in Section 7.



\section{Preliminaries}

\subsection{Nonhomogeneous linear recursion}

In this subsection, we firstly introduce the linear recurring sequence \cite{NLB}.

\textbf{Definition 1}. Let $k$ be a positive integer, and $c,a_{1},a_{2},\cdots,a_{k}$ be given elements of a finite field $GF(q)$ where $q$ is a prime. If $\{u_{i}\}_{i\geq 0}$ satisfies the relation
$$u_{i+k}=a_{k}u_{i+k-1}+\cdots+a_{1}u_{i}+c\quad(i=0,1,\cdots)\quad(\ast),$$
then $\{u_{i}\}_{i\geq 0}$ is called a $k$th-order linear recurring sequence in $GF(q)$.

\textbf{Remark 1}. Note that the terms $u_{0},u_{1},\cdots,u_{k-1}$, which can determine the rest of the sequence uniquely, are referred to as the initial values of the sequence. A relation of the form $(\ast)$ is called a $k$th-order linear recurrence relation. If $c$=0, we call $(\ast)$ a homogeneous linear recursion. Otherwise, $(\ast)$ is a nonhomogeneous linear recursion ($NLR$).

For a $k$th-order linear recurring sequence $\{u_{i}\}_{i\geq 0}$, $x^{k}+a_{1}x^{k-1}+\cdots+a_{k}=0$ is called its auxiliary equation, and $U(x)=\Sigma_{i=0}^{\infty}u_{i}x^{i}$ is called its generating function.

\textbf{Lemma 1}. Let $GF(q)$ be a finite field, where $q$ is a prime. Suppose that $(x-\alpha_{1})^{m_{1}}(x-\alpha_{2})^{m_{2}}\cdots(x-\alpha_{l})^{m_{l}}=0$ is the auxiliary equation of a $k$th-order linear recurring sequence $\{u_{i}\}_{i\geq 0}$ in $GF(q)$, where $m_{1}+m_{2}+\cdots+m_{l}=k$. Then the generating function of $\{u_{i}\}_{i\geq 0}$ is $$U(x)=\frac{R(x)}{(1-\alpha_{1}x)^{m_{1}}(1-\alpha_{2}x)^{m_{2}}\cdots(1-\alpha_{l}x)^{m_{l}}},$$
where $R(x)$ is a polynomial of $x$ with $\deg(R(x))<k$ in $GF(q)[X]$.

Further, $u_{i}=p_{1}(i)\alpha_{1}^{i}+p_{2}(i)\alpha_{2}^{i}+\cdots+p_{l}(i)\alpha_{l}^{i}$, where $p_{j}(i)=A_{0}+A_{1}i+A_{2}i^{2}+\cdots+A_{m_{j-1}}i^{m_{j-1}}$, $j=1,2,\cdots,l$. Notice that $A_{0},A_{1},\cdots,A_{m_{j-1}}$ are undetermined constants in $GF(q)$ which can be calculated from $a_{1},a_{2},\cdots,a_{k}$.

\textbf{Corollary 1}. Let $GF(q)$ be a finite field, where $q$ is a prime. Consider a typical fraction $\dfrac{R(x)}{(1-\alpha x)^{m}}$, where $\alpha \in GF(q)$, and $R(x)$ is a polynomial of $x$ with $\deg(R(x))<m$ in $GF(q)[X]$. Then,
$$\dfrac{R(x)}{(1-\alpha x)^{m}}=\sum_{i=0}^{\infty}u_{i}x^{i}$$
and $u_{i}=p(i)\alpha^{i}$, where $p(i)=A_{0}+A_{1}i+\cdots+A_{m-1}i^{m-1}$ and $A_{0},A_{1},\cdots,A_{m-1}$ are in $GF(q)$.

Through Corollary 1, the following two main theorems are given, which have been proved in \cite{DM2}.

\textbf{Theorem 1}. Let $GF(q)$ be a finite field, where $q$ is a prime. Assume that the sequence $(u_{i})_{i\geq 0}$ is defined by the following $NLR$ equations:
\begin{equation}
[NLR1]=\left\{
\begin{aligned}
& u_{0}=c_{0},u_{1}=c_{1},\cdots,u_{k-1}=c_{k-1},\\
&\sum_{j=0}^{k}\left( {\begin{array}{*{20}{ccc}}
	k\\
	j
	\end{array}} \right)u_{i+k-j}=c(-1)^{i}i \quad (i\geq 0),
\end{aligned}
\right.
\end{equation}
where $c,c_{0},c_{1},\cdots,c_{k-1}$ are constants in $GF(q)$. Therefore, $u_{i}=p(i)(-1)^{i}$, where $p(i)=A_{0}+A_{1}i+\cdots+A_{k+1}i^{k+1}$ and $A_{0},A_{1},\cdots,A_{k+1}$ are in $GF(q)$.

\textbf{Theorem 2}. Let $GF(q)$ be a finite field, where $q$ is a prime. Assume that the sequence $(u_{i})_{i\geq 0}$ is defined by the following $NLR$ equations:
\begin{equation}
[NLR2]=\left\{
\begin{aligned}
& u_{0}=c_{0},u_{1}=c_{1},\cdots,u_{k-1}=c_{k-1},\\
&\sum_{j=0}^{k}\left( {\begin{array}{*{20}{ccc}}
	k\\
	j
	\end{array}} \right)(-1)^{j}u_{i+k-j}=ci \quad (i\geq 0),
\end{aligned}
\right.
\end{equation}
where $c,c_{0},c_{1},\cdots, c_{k-1}$  are constants in $GF(q)$. Therefore, $u_{i}=p(i)$, where $p(i)=A_{0}+A_{1}i+\cdots+A_{k+1}i^{k+1}$, and $A_{0},A_{1},\cdots,A_{k+1}$ are in $GF(q)$.




\subsection{The XTR public key system}

In 2000, Lenstra and Verheul proposed the XTR public key system \cite{XTR}, utilizing the third-order LFSR sequence. Actually, XTR public key system belongs to LFSR sequence public key system\cite{GH1,GH2,XTR}.

Firstly, in 1999, Gong and Harn proposed LFSR public key cryptosystem \cite{GH1,GH2}, i.e., GH public key system, which is based on a third-order LFSR sequence generated by an irreducible polynomial $f(x)=x^{3}-ax^{2}+bx-1$, where $a,b\in GF(p)$ and $p$ is a prime.

Compared with LFSR public key cryptosystem, XTR public key system requires $b=a^{p}$ where $a\in GF(p^{2})$, and a special group with order $q$ in $GF(p^{6})^{*}$ where $q|(p^{2}-p+1)$ and $q>3$. In other words, the irreducible polynomial used in XTR public key system is $f(x)=x^{3}-ax^{2}+a^{p}x-1$, where $a \in GF(p^{2})$. Actually, XTR is the first method which utilizes computations on $GF(p^{2})$ to achieve $GF(p^{6})$ security without requiring explicit construction of $GF(p^{6})$.

Then we review some basic knowledge about the XTR public key system, which can be found in \cite{XTR,XTR2}.

\textbf{Definition 2}. Let $p>3$ and $q>3$ be two primes, satisfying $p\equiv 2\pmod{3}$, and $q|(p^{2}-p+1)$. Let $g$ be an element with order $q$ in $GF(p^{6})^{\ast}$, where $GF(p^{6})^{\ast}$ is the multiplicative group of the finite field $GF(p^{6})$. Then, we refer to the subgroup $<g>$ as the XTR group.

\textbf{Definition 3}. The conjugates of $g\in GF(p^{6})^{*}$ over $GF(p^{2})$  are $g, g^{p^{2}}$ and $g^{p^{4}}$. Then the trace function $Tr(g)$ of $g\in GF(p^{6})^{*}$ over $GF(p^{2})$ is the sum of the conjugates of $g$ over $GF(p^{2})$, which means that
$$Tr(g)=g+g^{p^{2}}+g^{p^{4}}.$$

\textbf{Proposition 1}. We have $Tr(g)^{p^{2}}=Tr(g)$, so that $Tr(g)\in GF(p^{2})$.

\textbf{Proposition 2}. The conjugates of $g$ with order $q$ satisfying $q|(p^{2}-p+1)$ are $g, g^{p-1}$ and $g^{-p}$, then we have $Tr(g)=g+g^{p-1}+g^{-p}\in GF(p^{2})$.

\textbf{Lemma 2}. The roots of $X^{3}-Tr(g)X^{2}+Tr(g)^{p}X-1$ are the conjugates of $g$, where $p$ is a prime.

By trace function, we can not only represent the elements of the XTR group by elements in $GF(p^{2})$, but also compute the powers of $g$ efficiently by performing the computations on $GF(p^{2})$ while avoiding operations in $GF(p^{6})$.


\textbf{Definition 4}. Let $c=Tr(g)\in GF(p^{2})$, we define
$$F(c,X)=X^{3}-cX^{2}+c^{p}X-1,$$
which is a polynomial in $GF(p^{2})[X]$. Let $g_{0},g_{1},g_{2}\in GF(p^{6})$ be three roots of $F(c,X)$, and for an integer $n\in Z$, we define $c_{n}=g_{0}^{n}+g_{1}^{n}+g_{2}^{n}$, which means that $c_{n}=Tr(g^{n})$. Obviously, $c=c_{1}$.

\textbf{Theorem 3}. $F(c,X)\in GF(p^{2})[X]$ is irreducible if and only if its roots have order $q$, where $q|(p^{2}-p+1)$ and $q>3$.

Next, we introduce the definition of the XTR-discrete logarithm (XTR-DL) problem.

\textbf{Definition 5}. Given $c=Tr(g)$, $c_{n} \in Tr(<g>)$, the XTR-DL problem is to find $0\leq n< q$ such that $c_{n}=Tr(g^{n})$.

The following theorem has been proved in \cite{XTR,XTR2}.

\textbf{Theorem 4}. The XTR-DL problem is equivalent to the discrete logarithm problem in $<g>$.

The security of XTR public key system is based on constructing a one-way trapdoor function through XTR-DL problem. In order to achieve this goal, when we know the values of $Tr(g)$ and $n$, we need to compute the values of $Tr(g^{n})$ efficiently, which has been solved by Lenstra and Verheul in \cite{XTR}.

Finally, we give the definition of XTR public key system.

\textbf{Definition 6}. Let $p>3$ and $q>3$ be two primes such that $p\equiv 2\pmod{3}$ and $q|(p^{2}-p+1)$. Let $g$ be an element in $GF(p^{6})^{*}$ with order $q$, and $\{p,q,g,Tr(g)\}$ be public parameters. All the computations here are implemented in $GF(p^{2})$:

(1) Public key: $Tr(g^{k})$, where $1<k<q$.

(2) Secret key: $k$, where $1<k<q$.

(3) Encryption: Given the plaintext $M \in GF(p^{2})$ and a secret random integer $b$ $(1<b<q-2)$, the ciphertext is $c=e_{k}(M,b)=(Tr(g^{b}),E)$, where $E=Tr(g^{bk})\ast M$, and $Tr(g^{bk})$ can be computed by Algorithm 2.3.7 \cite{XTR} using $b$ and $Tr(g^{k})$.

(4) Decryption: Given the ciphertext $c=(Tr(g^{b}),E)$, and the secret key $k$, and $Tr(g^{bk})$ can be computed by Algorithm 2.3.7 \cite{XTR} using $k$ and $Tr(g^{b})$. Then, the plaintext is $M=E\ast Tr(g^{bk})^{-1}$.

\subsection{Review and attack on DM2 schemes}
In this subsection, we review DM2 schemes \cite{DM2} simply which are based on ECC, and then provide a kind of attack on DM2 schemes. Because the two schemes in \cite{DM2} are similar, we take the type 1 scheme as an example.

\subsubsection{Review of DM2 schemes}

\textbf{Initialization phase}

Let $S_{1},S_{2},\cdots,S_{l}$ be $l$ shared secrets among $m$ participants $P_{1},P_{2},\cdots,P_{m}$. Let $q$ be a prime number such that $q>\left( {\begin{array}{*{20}{ccc}}
	k\\
	j
	\end{array}} \right)$ for $j=1,2,\cdots,k$, where $k$ is the threshold of this scheme.

Firstly, the dealer $D$ chooses two large primes $p_{1}$ and $p_{2}$, and calculates $N=p_{1}p_{2}$. Let $v$ be an integer such that $gcd(27v^{2},N)=1$. The elliptic curve $E_{N}(0,v)$ over the ring $\mathbb{Z}_{N}$ is the set of points $(x,y)\in \mathbb{Z}_{N} \times \mathbb{Z}_{N}$ satisfying the equation $y^{2}\equiv x^{3}+v \pmod{N}$ together with the point at infinity $\mathcal{O}_{N}$.

Then, $D$ considers $Q \in E_{N}(0,v)$ such that the discrete logarithm problem is infeasible in cyclic group $<Q>$. Finally, $D$ publishes $\{N,Q\}$.

Every participant $P_{i}$ chooses an integer $s_{i}$ randomly as secret shadow and calculates $R_{i}=s_{i}Q$. Then $P_{i}$ sends $(R_{i},i)$ to the dealer $D$. $D$ must ensure that for all $i\neq j$ , $R_{i}\neq R_{j}$. Finally, $D$ releases $(R_{1},R_{2},\cdots,R_{m})$.

\noindent\textbf{Construction phase}

The following steps need to be performed by $D$:

(1) $D$ chooses a random integer $e$ such that $gcd(e,n_{N})=1$ and calculates $d$ such that $ed\equiv de\equiv 1\pmod{n_{N}}$, where $n_{N}=lcm(\# E_{p_{1}}(0,v),\# E_{p_{2}}(0,v))$, and $\# E_{p}(0,v)$ denotes the order (i.e., the number of points) of the elliptic curve $E_{p}(0,v)$.

(2) For $i=1,2,\cdots,m$, $D$ calculates $R_{0}=dQ$ and $B_{i}=dR_{i}$ over $E_{N}(0,v)$.

(3) For $i=1,2,\cdots,m$, $D$ calculates $I_{i}=x_{B_{i}}+y_{B_{i}}$, where $x_{B_{i}}$ and $y_{B_{i}}$ are the x-coordinate and the y-coordinate of the point $B_{i}$ over $E_{N}(0,v)$ respectively.

(4) $D$ chooses an integer $c$ $(c<q)$ and considers a NLR defined by the following equations:
\begin{equation}
\left\{
\begin{aligned}
& u_{0}=I_{1},u_{1}=I_{2},\cdots,u_{k-1}=I_{k},\\
&\sum_{j=0}^{k}\left( {\begin{array}{*{20}{ccc}}
	k\\
	j
	\end{array}} \right)(-1)^{j}u_{i+k-j}=ci \pmod{q}\quad(i\geq 0).
\end{aligned}
\right.
\end{equation}

(5) For $k\leq i\leq m+l+3$, $D$ calculates $u_{i}$.

(6) $D$ calculates $y_{i}=I_{i}-u_{i-1}$ for $k<i \leq m$ and $r_{i}=S_{i}-u_{m+i}$ for $1\leq i\leq l$.

(7) $D$ releases $(R_{0},e,r_{1},r_{2},\cdots,r_{l},y_{k+1},y_{k+2},\cdots,y_{m},$\\$u_{m+l+2},u_{m+l+3})$.

\noindent\textbf{Verification phase}

Every participants $P_{i}$ can compute $s_{i}R_{0}$ to obtain his or her share $B_{i}$ as follows:
$$s_{i}R_{0}=s_{i}dQ=ds_{i}Q=dR_{i}=B_{i}.$$

Assume that at least $k$ participants $\{P_{i}\}_{i=1}^{k}$ use their shares $\{B_{i}\}_{i=1}^{k}$ to recover the secrets $S_{1},S_{2},\cdots,S_{l}$. A participant $P_{i}$ can check the validity of the secret shares provided by the other authorized participants by the steps as follows:

$eB_{j}=R_{j}$ over $E_{N}(0,v)$ for $j=1,2,\cdots,k$ and $j\neq i$.

\noindent\textbf{Recovery phase}

Assume that any $k$ participants $\{P_{i}\}_{i\in I}$ use their shares $\{B_{i}\}_{i\in I}$ to recover the secrets:

(1) Calculate $I_{i}=x_{B_{i}}+y_{B_{i}}$ for $i\in I$.

(2) Calculate $k$ terms $\{u_{i-1}\}_{i\in I}$ in the equations (3) using the formulas as follows:
$$
u_{i-1}=\left\{
\begin{aligned}
& I_{i}\qquad\qquad\qquad if\:1\leq i\leq k,\\
& I_{i}-y_{i}\qquad\qquad if\:k< i\leq m.
\end{aligned}
\right.
$$

(3) Utilize $k+2$ pairs $(i-1,u_{i-1})_{i\in I}$, $(m+l+2,u_{m+l+2})$, and $(m+l+3,u_{m+l+3})$ to construct the polynomial $p(x)$ with degree $k+1$:
$$p(x)=\sum_{i\in I'}Y_{i}\prod_{j\in I',j\neq i}\frac{x-X_{j}}{X_{i}-X_{j}}\pmod{q},$$
$$\qquad\qquad\quad=A_{0}+A_{1}x+\cdots+A_{k+1}x^{k+1}\pmod{q}.$$
Notice we use $(X_{i},Y_{i})$ for $i\in I'$ where $I'=I\cup \{m+l+2,m+l+3\}$ to denote these $k+2$ pairs, respectively.

(4) Calculate $u_{j}=p(j)$ for $j=m+1,m+2,\cdots,m+l$.

(5) Recover $S_{j}=u_{m+j}+r_{j}$ for $j=1,2,\cdots,l$.

\subsubsection{Attack on DM2 schemes}
Notice that when authorized participants recover the secrets, these participants only check the validity of $B_{i}$ by whether $eB_{i}$ equals to $R_{i}$, while the consistence between $B_{i}$ and $\{u_{i}\}$ is not verified. Thus when the sequence $\{u_{i}\}$ or $\{y_{i}\}$ is generated in the construction phase, a malicious $D$ can substitute the true $B_{i}=dR_{i}$ with a fake $B_{i}'=dR_{i}'$ $(R_{i}' \neq R_{i})$ over $E_{N}(0,v)$, which means that:

(1) $D$ chooses a random integer $e$ such that $gcd(e,n_{N})=1$ and calculates $d$ such that $ed\equiv de\equiv 1\pmod{n_{N}}$.

(2) For $i=1,2,\cdots,m$, $D$ calculates $R_{0}=dQ$ and $B_{i}=dR_{i}$ over $E_{N}(0,v)$.

When $1\leq i\leq k$,

(3) $D$ replaces $B_{i}$ with $B_{i}'$ over $E_{N}(0,v)$ to calculate a new $I_{i}'=x_{B_{i}'}+y_{B_{i}'}$, where $x_{B_{i}'}$ and $y_{B_{i}'}$ are the x-coordinate and the y-coordinate of the point $B_{i}'$ respectively.

(4) $D$ selects an integer $c$ $(c<q)$ and considers the following formulas:
$$\left\{
\begin{aligned}
& u_{0}=I_{1},u_{1}=I_{2},\cdots,u_{i-1}=I_{i}',\cdots,u_{k-1}=I_{k},\\
& \sum_{j=0}^{k}\left( {\begin{array}{*{20}{ccc}}
	k\\
	j
	\end{array}} \right)(-1)^{j}u_{i+k-j}=ci\pmod{q} \quad (i\geq 0).
\end{aligned}
\right.$$
Then $D$ calculates $u_{i}$ for $k\leq i\leq m+l+3$.

(5) $D$ calculates $y_{i}=I_{i}-u_{i-1}$ for $k<i\leq m$, and $r_{i}=S_{i}-u_{m+i}$ for $1\leq i\leq l$.

(6) $D$ releases $(R_{0},e,r_{1},\cdots,r_{l},y_{k+1},y_{k+2},\cdots,y_{m},$\\$u_{m+l+2},u_{m+l+3})$.

When $k< i\leq m$,

(3') For $i=1,2,\cdots,m$, $D$ calculates $I_{i}=x_{B_{i}}+y_{B_{i}}$, where $x_{B_{i}}$ and $y_{B_{i}}$ are the x-coordinate and the y-coordinate of the point $B_{i}$ over $E_{N}(0,v)$ respectively.

(4') $D$ selects an integer $c$ $(c<q)$ and considers the following formulas:
$$\left\{
\begin{aligned}
& u_{0}=I_{1},u_{1}=I_{2},\cdots,u_{k-1}=I_{k},\\
& \sum_{j=0}^{k}\left( {\begin{array}{*{20}{ccc}}
	k\\
	j
	\end{array}} \right)(-1)^{j}u_{i+k-j}=ci\pmod{q} \quad (i\geq 0).
\end{aligned}
\right.$$
Then $D$ calculates $u_{i}$ for $k\leq i\leq m+l+3$.

(5') $D$ replaces the $I_{i}$ with $I_{i}'$ to calculate $y_{i}'=I_{i}'-u_{i-1}$, where $I_{i}'\neq I_{i}$, then calculates other $y_{j}=I_{j}-u_{j-1}$  $(k<j\leq m,j\neq i)$ and $r_{i}=S_{i}-u_{m+i}$ $(1\leq i \leq l)$ correctly.

(6') $D$ releases $(R_{0},e,r_{1},r_{2},\cdots,r_{l},y_{k+1},y_{k+2},\cdots,y_{i}',$\\$\cdots,y_{m},u_{m+l+2},u_{m+l+3})$.

In the recovery phase, since $P_{i}$ can not discover the replacement, $P_{i}$ still offers the true $B_{i}$ that conflicts with the sequence $\{u_{i}\}$ or $\{y_{i}\}$ produced by the dealer as above. So the recovered secrets are not valid. Nonetheless, at least $k$ participants without $P_{i}$ can reconstruct secrets successfully. Actually, it is difficult to verify which $I_{i}$ is substituted. So DM2 schemes \cite{DM2} cannot prevent this kind of malicious behavior of the dealer. In addition, if more than one $I_{i}$ is replaced by the dealer with some invalid $I_{i}'$, the circumstance is even more complex.

\section{Scheme 1}
In this section,  in order to get rid of the drawback as mentioned in Section 2.3.2, we propose a novel VMSS scheme by using $[NLR1]$, XTR public key system, discrete logarithm problem and XTR-DL problem.

\subsection{Initialization phase}
$D$ represents the dealer. Let $P=\{P_{1},P_{2},\cdots,P_{m}\}$ be the collection of participants, and $k$ $(k\leq m)$ be the threshold.

At first, the dealer $D$ performs the following operations:

(1) $D$ randomly chooses two primes $p,q$ $(p>3,q>3)$ with $\lambda$ bits satisfying $p\equiv 2\pmod{3}$, $q|(p^{2}-p+1)$ and
   $q>\left( {\begin{array}{*{20}{ccc}}
	k\\
	j
	\end{array}} \right)$
for $j=0,1,\cdots,k$.

(2) $D$ selects an element $g$ of $GF(p^{6})^{\ast}$ with order $q$ satisfying that XTR-DL problem with the base $g$ is infeasible. Then $D$ computes $Tr(g)$.

(3) $D$ chooses $b\in Z$ $(1< b< q-2)$ randomly, then computes $Tr(g^{b})$.

(4) $D$ releases $(\lambda,p,q,g,Tr(g),Tr(g^{b}))$.

Then, the authorized participants perform the following operations:

(1) Each $P_{i}$ with $ID_{i}$ chooses $x_{i}\in Z$ $(1<x_{i}< q)$ for $i=1,2,\cdots,m$.

(2) $P_{i}$ computes $y_{i}=Tr(g^{x_{i}})$ as his shadow for $i=1,2,\cdots,m$.

(3) $P_{i}$ provides $(ID_{i},y_{i})$ to $D$, and keeps $x_{i}$ secret, where $i=1,2,\cdots,m$.

$D$ must ensure that $y_{i}\neq y_{j}$ $(i\neq j)$, otherwise $P_{i}$ needs to select a new $x_{i}$ to recalculate $y_{i}$. Then $D$ releases $(ID_{i},y_{i})$ for $i=1,2,\cdots,m$.

\textbf{Remark 2}: After the initialization phase, all the public parameters can be reused. Since $D$ does not get useful information from participants' shadows, these shadows can also be reused.

\subsection{Construction phase}

Let $S_{1}, S_{2},\cdots, S_{l} \in GF(q)^{\ast}$ be $l$ secrets. Then $D$ generates a subshadow $u_{i}$ for each participant $P_{i}$ as follows:

(1) Randomly chooses $c_{i}\in GF(q)^{\ast}$ for $i=0,1,2,\cdots,k-1$.

(2) Chooses a random constant $c\in GF(q)^{\ast}$, considers $[NLR1]$ presented by the following equations and computes $u_{i}$ for $ k \leq i \leq m+l+1$:
$$[NLR1]=\left\{
\begin{aligned}
& u_{0}=c_{0},u_{1}=c_{1},\cdots,u_{k-1}=c_{k-1},\\
& \sum_{j=0}^{k}\left( {\begin{array}{*{20}{ccc}}
	k\\
	j
	\end{array}} \right)u_{i+k-j}=c(-1)^{i}i\pmod{q} (i\geq 0).
\end{aligned}
\right.$$

(3) Computes $z_{i} = S_{i} - u_{m+i-1}\pmod{q}$ for $i = 1, 2,\cdots, l$.

(4) Computes $Tr(g^{bx_{i}})$ by using $Tr(g^{x_{i}})$ and $b$, then $E_{i}=Tr(g^{bx_{i}})\ast u_{i-1}\pmod{q} $ for $1\leq i\leq m$.

(5) Computes $T_{i}=g^{u_{i-1}}\pmod{p^{2}}$ for $1\leq i\leq m$.

(6) Releases $(E_{1}, E_{2},\cdots, E_{m}, T_{1}, T_{2},\cdots, T_{m},z_{1},z_{2},\cdots,$\\$z_{l},c,u_{m+l},u_{m+l+1})$.

\subsection{Verification phase}
Each $P_{i}$ can get its subshadow $u_{i-1}$ by the following way. At first, $P_{i}$ can compute $Tr(g^{bx_{i}})$ by using $x_{i}$ and $Tr(g^{b})$ for $1\leq i\leq m$. Then $P_{i}$ will get $u_{i-1}$ by
$$u_{i-1}=E_{i}\ast Tr(g^{bx_{i}})^{-1}\pmod{q},\quad 1\leq i\leq m.$$

The validity and consistence of $P_{i}$'s subshadow $u_{i-1}$ with public messages can be checked as follows:
$$ \prod_{j=0}^{k}(T_{i+1+k-j})^{\left( {\begin{array}{*{20}{ccc}}
	k\\
	j
	\end{array}} \right)}\overset{?}{=}g^{c(-1)^{i}i}\pmod{p^{2}},$$

$$T_{i}\overset{?}{=}g^{u_{i-1}}\pmod{p^{2}}.$$

If the verification succeeds, $P_{i}$ thinks that its subshadow $u_{i-1}$ is true and is consistent with public messages. If every verification succeeds, participants think that $D$ is honest.

\subsection{Recovery phase}

Suppose that at least $k$ participants $\{P_{i}\}_{i\in I}$ $(I\subseteq\{1,2,\cdots,m\})$ use these subshadows $\{u_{i-1}\}_{i\in I}$ to recover the shared secrets. Every $P_{i}$ can check the validity of $\{u_{j-1}|j\in I, j\neq i\}$ as follows:
$$g^{u_{j-1}}\overset{?}{=}T_{j}\pmod{p^{2}},\quad j\in I \;and\; j\neq i.$$

There are two ways to recover the secrets. From these two ways, we can see that Scheme 1 is a $(k,l,m)$-threshold secret sharing schemes.

Way 1: Owning $k$ true subshadows $\{u_{i-1}| i\in J \subseteq I, |J|=k\}$ and the published $\{u_{m+l},u_{m+l+1}\}$, they can use Theorem 1 to get the following equations, where $i\in J'=J \cup \{m+l+1, m+l+2\}$:
$$z_{0}+z_{1}(i-1)+\cdots+z_{k+1}(i-1)^{k+1}=u_{i-1}(-1)^{i-1}\pmod{q}.$$
Solving these $k+2$ equations or using Lagrange interpolation formulas, they have $z_{0} = A_{0}, z_{1} =
A_{1},\cdots,z_{k+1} = A_{k+1}$ in $GF(q)$.

Next, they get
\begin{small}
$$u_{i-1}=(A_{0}+A_{1}(i-1)+\cdots+A_{k+1}(i-1)^{k+1})(-1)^{i-1}\pmod{q}$$
\end{small}
where $i\in\{1,2,\cdots,m+l+2\}\backslash J'$.

Finally, they recover the secrets: $S_{i}=z_{i}+u_{m+i-1}\pmod{q}$, $i=1,2,\cdots,l.$

Way 2: If owning $k$ successive $\{u_{i-1},u_{i},\cdots,u_{i+k-2}\}$, these participants can get $u_{j}$ $(j=i+k-1,i+k,\cdots,m+l-1)$ by the following equations:
$$\sum_{j=0}^{k}\left( {\begin{array}{*{20}{ccc}}
	k\\
	j
	\end{array}} \right)u_{n+k-j}=c(-1)^{n}n\pmod{q} \quad (n\geq 0).$$

Finally, they can recover the secrets: $S_{i}=z_{i}+u_{m+i-1}\pmod{q},\, i=1,2,\cdots,l.$

\section{Scheme 2}
In this section,  in order to get rid of the drawback as mentioned in Section 2.3.2, we propose a novel VMSS scheme by using $[NLR2]$, XTR public key system, discrete logarithm problem and XTR-DL problem.
\subsection{Initialization phase}
The initialization phase in Scheme 2 is the same as Scheme 1.
\subsection{Construction phase}
In this phase, we replace $[NLR1]$ with $[NLR2]$, and the rest is identical to Scheme 1.
$$[NLR2]=\left\{
\begin{aligned}
& u_{0}=c_{0},u_{1}=c_{1},\cdots,u_{k-1}=c_{k-1},\\
&\sum_{j=0}^{k}\left( {\begin{array}{*{20}{ccc}}
	k\\
	j
	\end{array}} \right)(-1)^{j}u_{i+k-j}=ci\pmod{q} (i\geq 0).
\end{aligned}
\right.$$

\subsection{Verification phase}
Each $P_{i}$ can get its subshadow $u_{i-1}$ by computing $u_{i-1}=E_{i}\ast Tr(g^{bx_{i}})^{-1}\pmod{q}$ for $1\leq i\leq m$. The validity and consistence of $P_{i}$'s subshadow $u_{i-1}$ with public messages can be checked as follows:
$$ \prod_{j=0}^{k}(T_{i+1+k-j})^{(-1)^{j}\left( {\begin{array}{*{20}{ccc}}
	k\\
	j
	\end{array}} \right)}\overset{?}{=}g^{ci}\pmod{p^{2}},$$

$$T_{i}\overset{?}{=}g^{u_{i-1}}\pmod{p^{2}}.$$

If the verification succeeds, $P_{i}$ thinks its subshadow $u_{i-1}$ is true and is consistent with public messages. If every verification succeeds, participants think that $D$ is honest.

\subsection{Recovery phase}
Assume that at least $k$ participants $\{P_{i}\}_{i\in I}$ $(I\subseteq \{1,2,\cdots,m\})$ use these subshadows $\{u_{i-1}\}_{i\in I}$ to recover the shared secrets. Each $P_{i}$ can check the validity of $\{u_{j-1}|j\in I, j\neq i\}$ as follows:

$$g^{u_{j-1}}\overset{?}{=}T_{j}\pmod{p^{2}},\quad j\in I \;and\; j\neq i.$$

There are two ways to recover the secrets. From these two ways, we can see that Scheme 2 is also a $(k,l,m)$-threshold secret sharing schemes.

Way 1: Owning $k$ true subshadows $\{u_{i-1}|i \in J \subseteq I, |J|=k\}$ and the published $\{u_{m+l},u_{m+l+1}\}$, they can use Theorem 2 to get the following equations, where $i\in J'=J\cup \{m+l+1,m+l+2\}$:
$$z_{0}+z_{1}(i-1)+\cdots+z_{k+1}(i-1)^{k+1}=u_{i-1}\pmod{q}.$$
Solving these $k+2$ equations or using Lagrange interpolation formulas, they get $z_{0} = A_{0}, z_{1} =A_{1},\cdots,z_{k+1} = A_{k+1}$ in $GF(q)$.

Next, they get
$$u_{i-1}=A_{0}+A_{1}(i-1)+\cdots+A_{k+1}(i-1)^{k+1}\pmod{q}$$
where $i\in\{1,2,\cdots,m+l+2\}\backslash J'$.

Finally, they recover the secrets: $S_{i}=z_{i}+u_{m+i-1}\pmod{q}$, $i=1,2,\cdots,l.$

Way 2: If owning $k$ successive $\{u_{i-1},u_{i},\cdots,u_{i+k-2}\}$, these participants can get $u_{j}$ $(j=i+k-1,i+k,\cdots,m+l-1)$ by the following equations:
$$\sum_{j=0}^{k}\left( {\begin{array}{*{20}{ccc}}
	k\\
	j
	\end{array}} \right)(-1)^{j}u_{n+k-j}=cn\pmod{q} \quad (n\geq 0).$$

Finally, they can recover these secrets: $S_{i}=z_{i}+u_{m+i-1}\pmod{q},\,i=1,2,\cdots,l.$

\section{Security analysis}
The security of presented schemes is based on the nonhomogeneous linear recursion, XTR public key system, discrete logarithm problem and XTR-DL problem. Then we analyze our schemes from three aspects.

\subsection{Correctness}
In this subsection, we discuss the correctness of our schemes.

\textbf{Theorem 5}. If the dealer and correlated participants behave honestly, any $k$ participants can reconstruct the shared secrets.

\textbf{Proof}. We can utilize two ways mentioned in Sections 3.4 and 4.4 to recover those secrets.

The correctness of Way 1 is based on solving $k+2$ simultaneous equations or using Lagrange interpolation polynomials with random $k$ subshadows $\{u_{i-1}|i\in J, |J|=k\}$ and the published $\{u_{m+l},u_{m+l+1}\}$. In Theorem 1 and Theorem 2, there are $k+2$ uncertain coefficients in $p(x)$. Therefore, $p(x)$ can be uniquely defined, which means that any authorized $k$ participants can reconstruct the secrets.

The correctness of Way 2 is based on a nonhomogeneous linear recursion with degree $k$. Note that we require that the indices of these subshadows are successive. Since $[NLR1]$ and $[NLR2]$ are both nonhomogeneous linear recursions with degree $k$, these participants have to compute $k$ terms $u_{j}$ $(j=i-1,i,\cdots,i+k-2)$ to obtain other $u_{j'}$ $(j'=i+k-1,i+k,\cdots,m+l-1)$, which means that they can recover the shared secrets.

\textbf{Remark 3}. Next, we will discuss the reason why we publish $\{u_{m+l},u_{m+l+1}\}$ instead of other subshadows.

At first, $\{u_{0},u_{1},\cdots,u_{m-1}\}$ are subshadows of participants $\{P_{1},P_{2},\cdots,P_{m}\}$ respectively. Besides, $\{u_{m},u_{m+1},\cdots,u_{m+l-1}\}$ are correlated to the shared secrets $\{S_{1},S_{2},\cdots,S_{l}\}$. Then only $u_{m+l}$ and $u_{m+l+1}$ not only can satisfy the requirement, but also will not disclose any information about subshadows and secrets.

\subsection{Verifiability}
\textbf{Theorem 6}. In the construction phase, it is impossible for the dealer to cheat participants.

\begin{figure}[!t]
\begin{center}
\begin{tikzpicture}[
  font=\sffamily,
  every matrix/.style={ampersand replacement=\&,column sep=2cm,row sep=2cm},
  source/.style={draw,thick,rounded corners,inner sep=.3cm},
  process/.style={draw,thick,circle,fill=green!20},
  sink/.style={source},
  datastore/.style={draw,very thick,shape=datastore,inner sep=.3cm},
  dots/.style={gray,scale=2},
  to/.style={->,>=stealth',shorten >=1pt,semithick,font=\sffamily\footnotesize},
  every node/.style={align=center}]

  \matrix{

    \node[sink] (1) {$P_{i}$};
      \& \node[source] (2) {$D$};\qquad\qquad\qquad
      \\\\
      \node[sink] (3) {$P_{i}$};
       \& \node[source] (4) {$D$};
     \\
  };

   \draw[to] (1) to[bend right=50]
      node[midway,below] {$R_{i}$\\$P_{i}:R_{i}=s_{i}Q\:,s_{i}\: is\: the \: secret \:shadow\: of P_{i}\:$\\
      $D: R_{0}=dQ, \: d \:is \: the \: secret\: shadow \: of \:D$\\
      $secret\: share: \: B_{i}=s_{i}R_{0}=dR_{i},\:I_{i}=x_{B_{i}}+y_{B_{i}}$\\
      $subshadow\: u_{i-1}=\begin{cases}
                  I_{i} ,\qquad 1\leq i\leq k\\
                  I_{i}-y_{i},k+1\leq i\leq m
                 \end{cases}$} (2);
  \draw[to] (2) to[bend right=50] node[midway,above] {DM2 schemes\\$R_{0}$}
      (1);
  \draw[to] (3) to[bend right=50]
      node[midway,below] {$y_{i}$\\$P_{i}:y_{i=}Tr(g^{x_{i}})$
      \\$D:E_{i}=Tr(g^{bx_{i}})\ast u_{i-1}$\\
      \; $u_{i-1}=E_{i}\ast Tr(g^{bx_{i}})^{-1}$\\$subshadow=u_{i-1},1\leq i \leq m $} (4);
  \draw[to] (4) to[bend right=50] node[midway,above] {our schemes\\$E_{i}$}
      (3);

\end{tikzpicture}

\end{center}
\captionsetup{justification=centering}
\caption{The difference between DM2 schemes and our schemes}\label{fig:figname}
\end{figure}
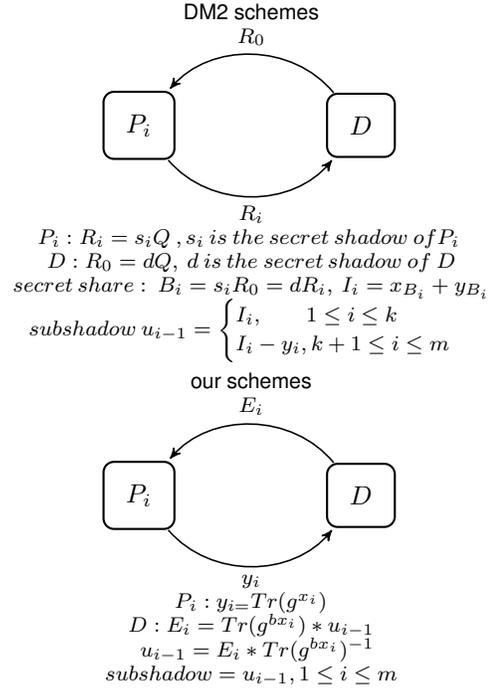

\textbf{Proof}. From the Figure 1, in DM2 schemes \cite{DM2}, we know that each $P_{i}$ chooses its secret shadow $s_{i}$, calculates $R_{i}=s_{i}Q$, and sends $R_{i}$ to $D$. After that, $D$ selects its secret shadow $d$ to compute $R_{0}=dQ$, then transforms $R_{0}$ to $P_{i}$. Hence, both $D$ and $P_{i}$ can calculate the secret share by $B_{i}=s_{i}R_{0}=dR_{i}$. Nevertheless, whether $B_{i}$ used in the generation of $\{u_{i-1}\}$ is identical to that offered by $P_{i}$ is not checked.

In contrast, in our schemes, $P_{i}$ chooses $x_{i}$ and keeps it secret from $D$. $P_{i}$ sends $y_{i}=Tr(g^{x_{i}})$ to the dealer, where $y_{i}$ is the public key of $x_{i}$. Then $D$ gets $E_{i}=Tr(g^{bx_{i}})\ast u_{i-1}$ by choosing a random number $b$ from $(1,q-2)$. After that, $P_{i}$ can compute $Tr(g^{bx_{i}})^{-1}$ by using its secret shadow $x_{i}$ and $Tr(g^{b})$. Finally, $P_{i}$ obtains its subshadow by $u_{i-1}=E_{i}\ast Tr(g^{bx_{i}})^{-1}$. Notice that we add consistence check between $u_{i-1}$ and public information so that the malicious dealer can be found.

We assume that the dealer can provide a false $P_{i}$'s subshadow $u_{i-1}'$ $(u_{i-1}'\neq u_{i-1})$ successfully in the construction phase, where $u_{i-1}$ is $P_{i}$'s valid subshadow. This implies that $T_{i}=g^{u_{i-1}}\equiv g^{u_{i-1}'}\pmod{p^{2}}$, and
$$ \prod_{j=0}^{k}(T_{i+1+k-j})^{\left( {\begin{array}{*{20}{ccc}}
	k\\
	j
	\end{array}} \right)}=g^{c(-1)^{i}i}\pmod{p^{2}}$$

or
$$ \prod_{j=0}^{k}(T_{i+1+k-j})^{(-1)^{j}\left( {\begin{array}{*{20}{ccc}}
	k\\
	j
	\end{array}} \right)}=g^{ci}\pmod{p^{2}}.$$
Because $u_{i-1},u_{i-1}'\in GF(q)$, the probability of $u_{i-1}'\neq u_{i-1}$ is negligible in the equations mentioned above, which means that it is impossible for the dealer to cheat participants successfully in the construction phase.

\textbf{Theorem 7}. In the recovery phase, it is impossible for the participant $P_{i}$ to cheat other participants and the dealer.

\textbf{Proof}. When a malicious $P_{i}$ provides $u_{i-1}'$ $(u_{i-1}'\neq u_{i-1})$ in the recovery phase, it implies that other participants can obtain $T_{i}=g^{u_{i-1}}\neq g^{u_{i-1}'}\pmod{p^{2}}$, which means that the malicious participant can be found in the recovery phase.

\textbf{Theorem 8}. In the recovery phase, it is impossible for two conspirators $P_{i}$ and $P_{j}$ to collude to cheat other participants and the dealer.

\textbf{Proof}. If $P_{i}$ and $P_{j}$ conspire, they exchange their secret key $x_{i}$ and $x_{j}$ privately. Thus $P_{i}$ gets $u_{j-1}$ and $P_{j}$ gets $u_{i-1}$, which means that they can pass through the verification phase. Nevertheless, all the participants have transmitted $(ID_{i},y_{i})$ to the dealer $D$ and $D$ has released them in the initialization phase. In consequence, the published message pair can guarantee that other participants and the dealer will detect this conspiracy owing to the mismatch between $ID_{i}$ and $u_{j-1}$ or $ID_{j}$ and $u_{i-1}$ in the recovery phase.

\subsection{Privacy}
\textbf{Theorem 9}. We assume that discrete logarithm problem, XTR-DL problem with the base $g\in GF(p^{6})^{\ast}$ is infeasible, and XTR public key system is secure. Then, the adversary cannot get anything about the secrets and subshadows.

\textbf{Proof}. From the description of our schemes, the public messages are as follows.

(1) $(ID_{i},y_{i})$ for $i=1,2,\cdots,m$.

First, $D$ has released $y_{i}$ $(1\leq i\leq m)$. If participants $P_{j}$ $(j\neq i)$ or the dealer wants to derive secret key $x_{i}$ from $y_{i}=Tr(g^{x_{i}})$, which means that the XTR-DL problem can be solved, it is impossible under our assumption. The adversary cannot decrypt $u_{i-1}$ from $E_{i}$ without $x_{i}$, so they cannot recover the secrets.

(2)$E_{1},E_{2},\cdots,E_{m}$.

We have $E_{i}=Tr(g^{bx_{i}})\ast u_{i-1}$ $(1\leq i\leq m)$, where $E_{i}$ is the XTR encryption of $P_{i}$'s subshadow $u_{i-1}$. When the adversary want to obtain $u_{i-1}$ from $E_{i}$, they need to break XTR public key system. Therefore, the adversary cannot get any useful information of subshadows and secrets under the assumption.

(3)$T_{1},T_{2},\cdots,T_{m}$.

$T_{i}=g^{u_{i-1}}$ $(1\leq i\leq m)$ have been released in the construction phase. Because the security of our schemes is based on the intractability of discrete logarithm problem with the base $g$ in the finite field $GF(p^{6})^{\ast}$, it is impossible to get $u_{i-1}$ from $T_{i}$ under the assumption.

\textbf{Theorem 10}. Any $k-1$ or fewer participants cannot reconstruct these secrets.

\textbf{Proof}. We might consider the worst case. Assume that there are exactly $k-1$ participants $\{P_{i},P_{i+1},\cdots,P_{i+k-2}\}$, which means that they only have $k+1$ terms, i.e., subshadows $\{u_{i-1},u_{i},\cdots,u_{i+k-3}\}$ and the published $\{u_{m+l},u_{m+l+1}\}$. However, there are $k+2$ undefined coefficients of $p(x)$ mentioned in Theorem 1 and Theorem 2. Then we cannot determine the polynomial $p(x)$ uniquely. Therefore, it is impossible for them to recover other subshadows by using the corresponding nonhomogeneous linear recursions. Consequently, the secrets shared by the dealer cannot be derived from $k-1$ or fewer participants.

\section{Performance analysis}

\subsection{Public values}
At first, we compare other proposed schemes \cite{DM1,DM2,HLC,DM3,LZZ,YF} with our schemes from the perspective of the amount of public values, which is the main index to measure the efficiency of VMSS schemes.

In Table 1, we use abbreviation OS to represent our schemes. In order to compare these schemes, we assume that there are $m$ participants, $l$ shared secrets, and the threshold is $k$.

From Table 1, we know that LZZ, YF and our schemes need more public values than the other schemes. However, except these three schemes, the other presented schemes cannot resist some malicious behaviors of the dealer as mentioned in Section 2.3.2. Further, our schemes make use of XTR public key system, then we can achieve the same security level as LZZ and YF schemes with shorter key length. Therefore, our schemes are relatively efficient among all these schemes.

\begin{table*}[!htbp]
	\caption{\textbf{Comparison of the amount of the public values in proposed schemes}}
	\label{table}
	\setlength{\tabcolsep}{3pt} 
	\renewcommand\arraystretch{1.5} 

    \tablefirsthead{\hline\multicolumn{1}{|c|}{Scheme}&\multicolumn{2}{|c|}{Amount of public values}
     &\multicolumn{1}{|c|}{ Public values}\\\hline}
     \tablelasttail{\hline}

  \begin{supertabular}{|p{0.1\textwidth}|p{0.2\textwidth}|p{0.15\textwidth}|p{0.48\textwidth}|}
   \multirow{2}{*}{DM1\cite{DM1}} & \multicolumn{2}{c|} {Type1 \: $2(m+3)+l-k$ } & \makecell*[c]{$\{e,N,g,q,\alpha\},$\\$(r,\{G_{i}\}_{i=1}^{m},\{r_{i}\}_{i=1}^{l},\{y_{i}\}_{i=k+1}^{m})$ }\\
   \cline{2-4}
   &  \multicolumn{2}{c|}{Type2 \: $2(m+3)+l-k$ }&
   \makecell*[c]{$\{N,g,q,\alpha\},\:\{R_{i}\}_{i=1}^{m}$\\$(R_{0},f,\{r_{i}\}_{i=1}^{l},\{y_{i}\}_{i=k+1}^{m})$}\\
   \hline

  \multirow{1}{*}{DM2\cite{DM2}} & \multicolumn{2}{c|} {Type1\&2 \: $2(m+3)+l-k$} & \makecell*[c]{$\{N,Q\},\:\{R_{i}\}_{i=1}^{m}$\\
  $(R_{0},e,\{r_{i}\}_{i=1}^{l},\{y_{i}\}_{i=k+1}^{m}$,\\
  $u_{m+l+2},u_{m+l+3})$ }\\
   \hline

  \multirow{2}{*}{HLC\cite{HLC}} & \multicolumn{2}{c|} {Scheme1 \: $2m+l-k+6$} & \makecell*[c]{$\{N,a,b\},\{ID_{i},s_{e_{i}}(a,b)\}_{i=1}^{m}$\\
  $(s_{e_{0}}(a,b),s_{-e_{0}}(a,b),d)$,\\
  $(\{Y_{i}\}_{i=1}^{m},\{h(i)\}_{i=1}^{l-k})$ }\\
  \cline{2-4}
  &\multicolumn{2}{c|}{Scheme2 \: $2m+l-k+7$  } &
  \makecell*[c]{$\{N,a,b,\alpha,q_{1}\},\:\{ID_{i},s_{e_{i}}(a,b)\}_{i=1}^{m}$\\
  $(s_{e_{0}}(a,b),d,\{r_{i}\}_{i=1}^{l},\{Y_{i}\}_{i=k+1}^{m})$}\\
  \hline

  \multirow{1}{*}{DM3\cite{DM3}} & \multicolumn{2}{c|} {Type1\&2 \: $3m+l-k+7$} & \makecell*[c]{$\{N,a,b,q_{1}\},\{ID_{i},s_{e_{i}}(a,b)\}_{i=1}^{m},$\\
  $(s_{e_{0}}(a,b),d,\{r_{i}\}_{i=1}^{l},\{y_{i}\}_{i=k}^{m})$ }\\
  \hline

  \multirow{4}{*}{LZZ\cite{LZZ}} &  \multirow{2}{*}{Scheme1} &  $3m+k+5$, \: when $l\leq k$ & \makecell*[c]{$(\lambda,N,Q,q,g),\{ID_{i},e_{i},N_{i}\}_{i=1}^{m}$,\\
  $(\{C_{i}\}_{i=1}^{m},\{H_{i}\}_{i=1}^{m},\{A_{i}\}_{i=1}^{k})$}\\
  \cline{3-4}
  & & $3m+3l-2k+5$,\: when $l>k$ & \makecell*[c]{$(\lambda,N,Q,q,g),\{ID_{i},e_{i},N_{i}\}_{i=1}^{m}$,\\
  $(\{C_{i}\}_{i=1}^{m},\{H_{i}\}_{i=1}^{m},\{\eta_{i}\}_{i=1}^{l-k}$,\\
  $\{f(\eta_{i})\}_{i=1}^{l-k},\{A_{i}\}_{i=1}^{l})$}\\
  \cline{2-4}
  & \multicolumn{2}{c|}{Scheme2 \: $3m+l+6$}  &
  \makecell*[c]{$(\lambda,N,Q,q,g,\alpha),\:\{ID_{i},e_{i},N_{i}\}_{i=1}^{m}$\\
  $(\{H_{i}\}_{i=1}^{m},\{T_{i}\}_{i=1}^{m},\{Y_{i}\}_{i=1}^{l})$}\\
  \hline
   \multirow{1}{*}{YF\cite{YF}} & \multicolumn{2}{c|} {Scheme1\&2 \: $3m+l+7$} & \makecell*[c]{$(\lambda,N,Q,q,g),\:\{ID_{i},e_{i},N_{i}\}_{i=1}^{m}$\\
  $(\{H_{i}\}_{i=1}^{m},\{T_{i}\}_{i=1}^{m},\{y_{i}\}_{i=1}^{l}$,\\
  $c,u_{m+l})$ }\\
  \hline

  \multirow{1}{*}{OS} & \multicolumn{2}{c|} {Scheme1\&2 \: $3m+l+9$} & \makecell*[c]{$(\lambda,p,q,g,Tr(g),Tr(g^{b})),\:\{ID_{i},y_{i}\}_{i=1}^{m}$\\
  $(\{E_{i}\}_{i=1}^{m},\{T_{i}\}_{i=1}^{m},\{z_{i}\}_{i=1}^{l}$,\\
  $c,u_{m+l},u_{m+l+1})$}\\
  \hline
  \end{supertabular}
 \label{tab1}
\end{table*}

\subsection{Computational complexity}
Next, we compare the computational complexity of the presented schemes \cite{DM1,DM2,HLC,DM3,LZZ,YF} and our schemes. And we utilize the notations below in Table 2.

\begin{table}[!htbp]
	\caption{\textbf{Notations}}
	\label{table}
	\setlength{\tabcolsep}{2pt} 
	\renewcommand\arraystretch{1.5} 

    \tablefirsthead{\hline\multicolumn{1}{|c|}{Symbol}
    &\multicolumn{1}{|c|}{Explanation}\\\hline}
    \tablelasttail{\hline}

    \begin{supertabular}{|p{0.1\textwidth}|p{0.35\textwidth}|}

    \multirow{1}{*}{$T_{e}$}
    & \makecell*[c]{cost of one modular exponentiation \\ on some finite field}\\
    \hline
    \multirow{1}{*}{$T_{m}$}
    & \makecell*[c]{cost of one modular multiplication \\ on some finite field}\\
    \hline
    \multirow{1}{*}{$T_{L}(i)$}
    & \makecell*[c]{cost of the Lagrange basis of $i$ points, \\ where $i\geq 0$}\\
    \hline
    \multirow{1}{*}{$T_{M}$}
    & \makecell*[c]{average cost of a scalar multiplication \\ on the elliptic curve}\\
    \hline
    \multirow{1}{*}{$T_{le}(i)$}
    & \makecell*[c]{cost of obtaining a solution of $i$ linear equations, \\where $i\geq 0$}\\
    \hline
    \end{supertabular}
    \label{tab1}
\end{table}

Because all the schemes listed here are multi-use secret sharing schemes, which means that the initialization phase of every scheme needs to be performed only once, we ignore the cost of this phase in the following part. To save space, in Table 3, we use Con, Ver, Rec to represent the construction phase, verification phase, and recovery phase, respectively. Similarly, OS is on behalf of our schemes. In addition, we also assume that there are $m$ participants, $l$ shared secrets, and the threshold is $k$.

\begin{table*}
	\caption{\textbf{Comparison of the computational complexity in presented schemes}}
	\label{table}
	\setlength{\tabcolsep}{3pt} 
	\renewcommand\arraystretch{1.5} 

    \tablefirsthead{\hline\multicolumn{1}{|c|}{Scheme}
    &\multicolumn{1}{|c|}{ Con}&\multicolumn{1}{|c|}{ Ver}&\multicolumn{1}{|c|}{ Rec}\\\hline}
    \tablelasttail{\hline}

    \begin{supertabular}{|p{0.1\textwidth}|p{0.4\textwidth}|p{0.2\textwidth}|p{0.235\textwidth}|}

    \multirow{1}{*}{DM1\cite{DM1}}
   & \makecell*[c]{$T_{e}+kT_{m}$} &\makecell*[c]{$T_{e}$}& \makecell*[c]{$kT_{e}+T_{L}(k)$}\\

    \hline
    \multirow{1}{*}{DM2\cite{DM2}}
    & \makecell*[c]{$2T_{M}+(k+1)T_{m}$} &\makecell*[c]{$T_{M}$}& \makecell*[c]{Way1\: $T_{L}(k+2)$\\ Way2\: $T_{le}(k)$}\\

   \hline
   \multirow{2}{*}{HLC\cite{HLC}}
   & \makecell*[c]{Scheme1\\$3T_{e}+kT_{m}$\: ($l\leq k$)\\$3T_{e}+lT_{m}$\:$(l>k)$}
   & \makecell*[c]{$T_{e}$}
   &\makecell*[c]{Scheme1\\$T_{L}(k)$\: ($l\leq k$)\\$T_{L}(l)$\:$(l>k)$}\\
   \cline{2-2}
   \cline{4-4}
   & \multicolumn{1}{c|}{\makecell*[c]{Scheme2\\$3T_{e}+kT_{m}$}}
   &\makecell*[c]{}&\makecell*[c]{Scheme2\\$T_{L}(k)$}\\

   \hline
   \multirow{1}{*}{DM3\cite{DM3}}
   & \makecell*[c]{$3T_{e}+kT_{m}$ }
   &\makecell*[c]{$T_{e}$}&\makecell*[c]{Way1,2\: $T_{L}(k)$\\ Way3\: $T_{le}(k)$}\\

   \hline
    \multirow{2}{*}{LZZ\cite{LZZ}}
    & \makecell*[c]{Scheme1\\$2T_{e}+kT_{m}$\:$(l\geq k)$\\$2T_{e}+lT_{m}$\:$(l>k)$}
    & \makecell*[c]{Scheme1\\$T_{e}$\:$(l\leq k)$\\$2T_{e}$\:$(l>k)$}
    & \makecell*[c]{Scheme1\\$T_{L}(k)$\:$(l\leq k)$\\$T_{L}(l)$\:$(l>k)$}\\
   \cline{2-4}
    & \makecell*[c]{Scheme2\\$2T_{e}+kT_{m}$}
    & \makecell*[c]{Scheme2\\$2T_{e}$}
    &\makecell*[c]{Scheme2\\$T_{L}(k)\, or \, T_{le}(k)$}\\

   \hline
    \multirow{1}{*}{YF\cite{YF}}
    & \makecell*[c]{$2T_{e}+(k+1)T_{m}$}
    &\makecell*[c]{$2T_{e}$}
    &\makecell*[c]{Way1\: $T_{L}(k+1)$\\ Way2\: $T_{le}(k)$}\\

  \hline
   \multirow{1}{*}{OS}
   & \makecell*[c]{$T_{e}+(k+1)T_{m}$}
   &\makecell*[c]{$2T_{e}$}
   &\makecell*[c]{Way1\:$T_{L}(k+2)$\\ Way2\: $T_{le}(k)$}\\
   \hline
  \end{supertabular}
  \label{tab1}
\end{table*}

\subsubsection{Construction phase}
In the Scheme 1 of HLC and LZZ, they employ the polynomials of degree $k-1$ or $l-1$ to share secrets. However, DM1, DM2, DM3, YF, our schemes and the Scheme 2 of HLC, LZZ utilize the linear recursion. Because LZZ, YF and our schemes can detect the malicious behavior of the dealer, all these three schemes need more computations than the others.

The differences of computations among these three schemes lie in the different public key system used in them. Because the trace function used in the XTR public key system needs less time than modular multiplication computation used in RSA and LFSR public key cryptosystems, our schemes are faster to implement than LZZ and YF schemes in this phase.

What's more, the Scheme 1 of HLC and LZZ need two ways to deal with different cases, which are more complex to operate than the other schemes. Therefore, YF and our schemes are easier to run than LZZ.

\textbf{Remark 4}. Notice that we do not consider the cost of trace function in Table 3.

\subsubsection{Verification phase}
Except LZZ, YF and our schemes, the other schemes need less computations. Since these three schemes overcome the drawback mentioned before, they need more computations to verify the validity of shares. The serious consequences of the lack of these verification have been shown in Section 2.3.2.

\subsubsection{Recovery phase}
The recovery phase is the most time-consuming phase in these phases. In fact, all schemes mentioned here can use Lagrange interpolation polynomial to recover the shared secrets. However, DM1, DM2, DM3, YF, our schemes and the Scheme 2 of HLC, LZZ can make use of linear recursions to reconstruct secrets, which are much easier and faster to construct than Lagrange interpolation polynomial.

Because a polynomial of degree $n$ needs $O(n^{2})$  time to construct by Lagrange interpolation, the recovery phase of Scheme 1 of HLC and LZZ can be operated within $O(k^{2})$ $(l\leq k)$ or $O(l^{2})$ $(l>k)$ time.

YF and our schemes have two ways to recover the secrets. As for the Way 1, YF schemes need $O(k^{2})$ time, and our schemes need $O((k+1)^{2})$ time. Because we use different nonhomogeneous linear recursions from YF schemes, our schemes need more computations. The Way 2 is easier to implement than the first way, however it has stricter condition, which means that corresponding participants' indices must be consecutive. Since the nonhomogeneous linear recursions used in YF and our schemes are both $k$-th order, these two schemes need $k$ terms of subshadows to determine the nonhomogeneous linear recursions used in the construction phase, which means that they have the same computational complexity in the recovery phase for Way 2.

\textbf{Remark 5}. Compared with LZZ schemes using homogeneous linear recursions, YF and our schemes need more computations in Way 1, because these two schemes utilize nonhomogeneous linear recursions which are more complex than homogeneous linear recursions. Nevertheless, if we utilize the same linear recursion in these three schemes, they will cost the same time in the recovery phase for Way 1.

\subsection{Dynamic attribute}
Then, we will show a dynamic update, deletion, addition of the participants, the values of secrets and the threshold according to the actual situation.

\noindent{Participants:}

When a participants $P_{new}$ needs to be added in the scheme, $P_{new}$ selects an integer $x_{new}$ $(1<x_{new}< q)$ randomly and computes $y_{new}=Tr(g^{x_{new}})$, then transmits $(ID_{new}, y_{new})$ to the dealer. Next, $D$ can compute $E_{new}=Tr(g^{bx_{new}})\ast u_{new-1}\pmod{q}$ and $T_{new}=g^{u_{new-1}}\pmod{p^{2}}$ where $q|(p^{2}-p+1)$, and it releases them later. Similarly, when the scheme needs to delete a participant $P_{del}$, $D$ only erases $(ID_{del},y_{del})$ from its list. Therefore, if $P_{del}$ wants to attack the scheme by its subshadow $u_{del-1}$, the dealer will detect this malicious behavior.

\noindent{Secrets:}

Once $D$ wants to add a secret $S_{l+1}$ to the scheme, $D$ can obtain $z_{l+1}=S_{l+1}-u_{m+(l+1)-1}=S_{l+1}-u_{m+l}$. Likewise, $D$ can delete a secret $S_{i}$ by erasing $z_{i}=S_{i}-u_{m+i-1}$ from the list. If $D$ wants to update the secrets, $D$ only erases the old secrets and then add the new one into the scheme by corresponding operations mentioned above.

\noindent{Threshold:}

Our schemes are secure $(k,l,m)$-VMSS schemes, because our schemes make use of a $NLR$ of degree $k$. Therefore, if $D$ wants to change the threshold, $D$ can replace the original $NLR$ with a new degree, which means that our schemes are threshold changeable multi-secret sharing schemes.

\subsection{Performance feature}
\qquad Finally, we analyze performance features of the schemes in \cite{DM1,DM2,HLC,DM3,LZZ,YF} and our schemes in Table 4.

$\bullet$ Feature 1: Reconstruct multiple secrets at the same time

$\bullet$ Feature 2: Utilize the public channel

$\bullet$ Feature 3: Resist the conspiracy attack

$\bullet$ Feature 4: Update the secrets after an unsuccessful recovery

$\bullet$ Feature 5: Reuse the shadows with different access structure

$\bullet$ Feature 6: Reuse the shadows with different $D$

$\bullet$ Feature 7: Perceive $D$'s deception

$\bullet$ Feature 8: Perceive $P_{i}$'s deception

$\bullet$ Feature 9: The bit length of private key in a 1024-bit finite field

$\bullet$ Feature 10: The bit length of public key in a 1024-bit finite field

\begin{table*}[!htbp]\small
\caption{Performance\, feature}
\label{tab:1}
\scalebox{1.4}[1.1]{
\begin{tabular}{|c|c|c|c|c|c|c|c|}
\hline
Feature  & DM1\cite{DM1} & DM2\cite{DM2} & HLC\cite{HLC} &  DM3\cite{DM3} & LZZ\cite{LZZ} & YF\cite{YF} & OS\\
\hline
1  & YES & YES & YES & YES & YES & YES & YES\\
\hline
2 & YES & YES & YES & YES & YES & YES & YES\\
\hline
3 & NO & NO & YES & YES & YES & YES & YES\\
\hline
4 & NO & NO & NO & NO & NO & NO & NO\\
\hline
5 & YES & YES & YES & YES & YES & YES & YES\\
\hline
6 & YES & YES & YES & YES & YES & YES & YES\\
\hline
7 & NO & NO & NO & NO & YES & YES & YES\\
\hline
8 & YES & YES & YES & YES & YES & YES & YES\\
\hline
9 & 1024 & 1024 & 340 & 340 & 1024 & 340 & 170 \\
\hline
10 & 1024 & 1024 & 340 & 340 & 1024 & 340 & 340 \\
\hline
\end{tabular}}
\end{table*}

From Table 4, we know that DM1 and DM2 schemes cannot resist conspiracy attack as analyzed in Theorem 8, because the dealer does not construct the links between the identity messages and corresponding secret shadows of the specific participants.

Notice that, except LZZ, YF and our schemes, the other schemes cannot perceive malicious dealer, since they lack verification between their participants' subshadows and public messages.

Nevertheless, in a 1024-bit finite field, the length of our private key can be one-sixth of LZZ schemes, and one-third of YF schemes, which is about 170 bits. And the length of the public key can be one-third of LZZ schemes, and equal to YF schemes, which is about 340 bits. This is because the private key $x_{i}$ is in $GF(q)$, and the public key $Tr(g^{x_{i}})$ is in $GF(p^{2})$. What's more, the XTR public key system used in our schemes can realize the security level in $GF(p^{6})$ by computations in $GF(p^{2})$. It has been proved that the security level of a 170-bit XTR is equivalent to a 340-bit LFSR public key cryptosystem or a 1024-bit RSA public key cryptosystem, which means that our schemes can achieve the same security level as LZZ and YF schemes with shorter key size.

Therefore, our schemes are better schemes than the other schemes mentioned in this section.

\textbf{Remark 6}. If we can use LFSR sequences with a higher order, such as sixth-order, to construct a new LRSR sequence public key system applied to our schemes, then we will use shorter key size. However, there is not necessarily a fast way to get the required parameters in the whole scheme. So we choose XTR public key system generated by a third-order LFSR sequence to construct our new VMSS schemes.

\section{Conclusion}

\qquad In this paper, we utilize XTR public key system to construct two new efficient VMSS schemes which are improved versions of the VMSS schemes proposed by Dehkordi and Mashhadi in 2008.

Compared with the previous presented schemes, our schemes can detect the malicious dealer by adding verification between participants' subshadows and public messages. Even though LZZ and YF schemes have the same advantages as our schemes, we use shorter key size to achieve the same security level. In addition, our schemes are efficient to implement because they have dynamic attributes, which means that our schemes can change the number of participants, the values of secrets and the threshold easily according to the practical situation.

In conclusion, our schemes are computationally secure $(k,l,m)$-VMSS schemes which can share multiple secrets simultaneously, use the public channel, have verifiability, reuse subshadows, and are both dynamic and threshold changeable with shorter parameters.


%



\ifCLASSOPTIONcompsoc
  \section*{Acknowledgments}
\else
  \section*{Acknowledgment}
\fi
This research is supported by the National Key Research and Development Program of China (Grant No. 2018YFA0704703), the National Natural Science Foundation of China (Grant No. 61971243), the Fundamental Research Funds for the Central Universities of China, and the Nankai Zhide Foundation.

\ifCLASSOPTIONcaptionsoff
  \newpage
\fi



%

%

\begin{IEEEbiographynophoto}{Jing Yang}
received the B.S. degree in mathematics and applied mathematics from Shandong Normal University, Jinan, China in 2015, and M.S. degree in Probability and Mathematical Statistics from Nankai University,
Tianjin, China in 2019. She is currently a Ph.D. student advised by Prof. Fang-Wei Fu in Chern Institute of Mathematics and LPMC, and Tianjin Key Laboratory of Network and Data Security Technology, Nankai University, Tianjin, China. Her research interests include secret sharing, blockchain, and corresponding cryptography.
\end{IEEEbiographynophoto}

\begin{IEEEbiographynophoto}{Fang-Wei Fu}
received the B. S. degree in mathematics, the M. S. degree, and the Ph.D. degree in applied mathematics from Nankai University, Tianjin, China, in 1984, 1987 and 1990, respectively. Since April 2007, he has been with the Chern Institute of Mathematics, Nankai University, Tianjin, China, where he is a Professor. From June 1987 to April 2007, he was with the School of Mathematical Science, Nankai University, Tianjin, China, and became a Professor there in 1995. From February 2002 to March 2007, he was a Research Scientist with the Temasek Laboratories, National University of Singapore, Republic of Singapore. From November 1989 to November 1990, he visited the Department of Mathematics, University of Bielefeld, Germany. From October 1996 to October 1997, he visited the Institute for Experimental Mathematics, University of Duisburg-Essen, Germany. He also visited the Department of Information Engineering, The Chinese University of Hong Kong, Hong Kong, the Department of Mathematics, University of California, Irvine, USA, the Division of Mathematical Sciences, the School of Physical and Mathematical Sciences, Nanyang Technological University, Republic of Singapore. His current research interests include coding theory, cryptography, and information theory.
\end{IEEEbiographynophoto}





\end{document}